\documentclass[aip,jap,noshowpacs,noshowkeys,reprint]{revtex4-1}
\usepackage{amsmath}
\usepackage{graphicx}

\begin{document}

\title{A model for pressurized hydrogen induced thin film blisters} 

\author{R.A.J.M. van den Bos}
\email{r.a.j.m.vandenbos@utwente.nl}
\affiliation{Industrial Focus Group XUV Optics, MESA+ Institute for Nanotechnology, University of Twente, Enschede, The Netherlands}

\author{V. Reshetniak}
\affiliation{Institute for Spectroscopy Russian Academy of Sciences (ISAN), Troitsk, Moscow, Russian Federation}

\author{C.J. Lee}
\affiliation{Industrial Focus Group XUV Optics, MESA+ Institute for Nanotechnology, University of Twente, Enschede, The Netherlands}

\author{J. Benschop}
\affiliation{Industrial Focus Group XUV Optics, MESA+ Institute for Nanotechnology, University of Twente, Enschede, The Netherlands}
\affiliation{ASML Netherlands B.V., Veldhoven, The Netherlands} 

\author{F. Bijkerk}
\affiliation{Industrial Focus Group XUV Optics, MESA+ Institute for Nanotechnology, University of Twente, Enschede, The Netherlands}

\date{\today}

\begin{abstract}
We introduce a model for hydrogen induced blister formation in nanometer thick thin films. The model assumes that molecular hydrogen gets trapped under a circular blister cap causing it to deflect elastically outward until a stable blister is formed. In the first part, the energy balance required for a stable blister is calculated. From this model, the adhesion energy of the blister cap, the internal pressure and the critical H-dose for blister formation can be calculated. In the second part, the flux balance required for a blister to grow to a stable size is calculated. The model is applied to blisters formed in a Mo/Si multilayer after being exposed to hydrogen ions. From the model the adhesion energy of the Mo/Si blister cap was calculated to be around 1.05 J/m$^2$ with internal pressures in the range of 175-280 MPa. Based on the model a minimum ion dose for the onset of blister formation was calculated to be $d=4.2\times10^{18}$  ions/cm$^2$. From the flux balance equations the diffusion constant for the Mo/Si blister cap was estimated to be $D_{H_2}=(10\pm1)\times10^{-18}$ cm$^2$/s.
\end{abstract}


\maketitle

\section{\label{sec:intro}introduction}
Nanometer thick multilayer structures can be designed and fabricated to form an artificial Bragg structure that can be used to reflect light of a specific wavelength. These mirrors can be found in synchrotrons, telescopes and extreme ultraviolet optical systems \cite{pelizzo2011,teyssier2002}. Over time, the surfaces of such mirrors can become contaminated, with a loss in reflectance as a result. One way to remove these contaminants from the surface is by exposing the surface to a flux of ionic and/or atomic hydrogen. However, under certain hydrogen exposure conditions, surface blisters may appear which irreversibly damage the mirror surface. 

Blister formation is not exclusively related to multilayer mirrors but can also be found in a much broader research field. For example in fusion reactor wall studies and the smart-cut process for silicon on insulator fabrication \cite{evans1976,feng2004,hochbauer2005,sharafat2009,xie2015}. Blisters have been observed in both heterogeneous nanometer thick layered structures, and also in bulk materials. In addition to hydrogen, helium ions have been found to induce blistering \cite{johnson1999}. Based on the experimentally observed blisters, several models have been developed to predict the critical dose for the onset of blister formation, adhesion energy and radius of the blisters \cite{freund1997,hong2006,mitani1992,selvadurai2007}. In general these models are based on calculating the potential energy of the blister cap as a function of pressure, volume and elastic constants of the cap material. When the strain energy of deformation plus the surface energy is balanced by the mechanical work of the gas trapped inside the blister a stable blister cap is formed.

\begin{figure}
	\includegraphics[width=1\columnwidth]{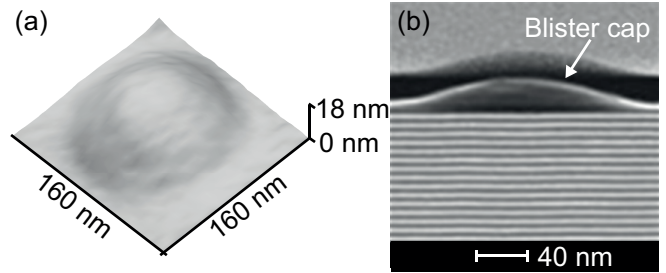}
	\caption{\label{fig:TEMimage}Two examples of surface blisters formed on a Mo/Si multilayer after exposure to 200 eV hydrogen ions. AFM image (a) and a cross sectional TEM image (b). The TEM image shows a delamination of the first Mo/Si bilayer (Mo bright, Si dark).}
\end{figure}

In this article, a blister formation model, based on pressure driven elastic deformation, is introduced. Special attention is paid to blisters formed in a Mo/Si multilayer by hydrogen ions, of which examples are shown in figure \ref{fig:TEMimage}. Here, we extend the previously described models in the following way. The compressive stress introduced during deposition is taken into account and in place of the ideal gas law an empirical equation of state (EOS), suitable for high pressure is used. Furthermore, we use the stable blister size to estimate the diffusion of molecular hydrogen through the blister cap. We show that the model agrees with experimental data. Finally, the model predictions for the influence of initial intrinsic stress, the adhesion energy, the blister’s internal pressure, and a minimum hydrogen dose for the onset of blisters are discussed.

\section{\label{sec:theory}theory}
Blister formation is a multi-step mechanism that can qualitatively be described by the following steps: i) Atomic and/or ionic hydrogen penetrates into the substrate; ii) Because the solubility of hydrogen in the target material is limited, hydrogen segregates into micro cavities and defect sites where it can recombine to molecular hydrogen and gets trapped; iii) The pressure inside the cavity increases as more molecular hydrogen is accumulated up to the point a blister is formed iv) The blister either stops growing or bursts depending on the transport of hydrogen through the material \cite{terreault2007}.

To calculate the blister's energy balance, a blister shape must be assumed. Under the assumption that the blister cap can be described as an isotropic elastic thin film that deflects due to the pressure in the blister cavity a stable blister size can be calculated depending on the number of trapped molecular hydrogen particles . 

\subsection{\label{subsec:blistershape}Blister shape function}
\begin{figure}
	\includegraphics[width=1\columnwidth]{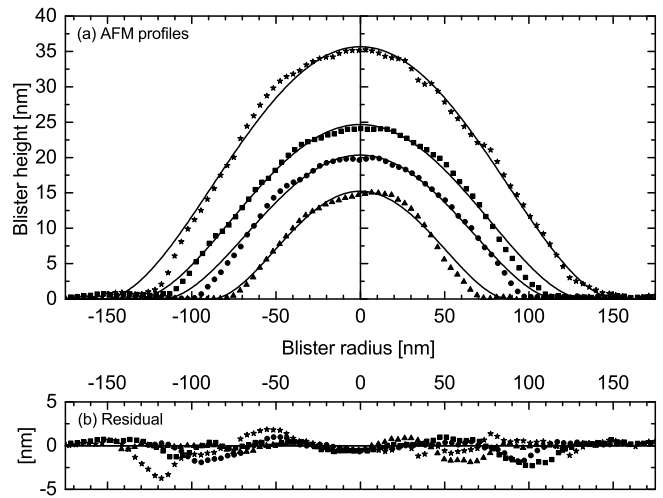}
	\caption{\label{fig:AFMprofile}Measured profiles of four blister caps (points) fitted with analytical expression of equation \ref{eq:blistershape} (solid lines). Bottom graph (b) gives residual of fitted function.}
\end{figure}

A commonly used function for describing the blister shape is a bell shaped profile function as given by \cite{hong2006,selvadurai2007,coupeau2013}:
\begin{equation}
z(r)=\left\{ 
\begin{array}{lr}
z_0\left(1-\left(\frac{r}{r_0}\right)^2\right)^2 & r\leq r_0.\\
0 & r>r_0.
\end{array}\right.
\label{eq:blistershape}
\end{equation}
In this formula, $z(r)$ is the height of the blister cap at a distance $r$ from the blister top, which has a deflection $z_0$. The blister radius is given by $r_0$. This function is a solution of the classical plate equation from Poisson-Kirchhoff-Germain thin plate theory for small deflections \cite{timoshenko1987}:
\begin{equation}
D\nabla^2\nabla^2z(r)=p,
\label{eq:plateeqn}
\end{equation} 
with $p$ the pressure inside the blister and\newline $D=Et^3 (12(1-\nu^2))^{-1}$ the plate constant. The plate constant is determined by the blister cap thickness $t$, Young’s modulus, $E$, and Poisson’s ratio, $\nu$. For a circular plate with fixed boundaries, i.e.
\begin{equation}
\begin{array}{cc}
2D\frac{1}{r}\frac{d}{dr}\left[\frac{1}{r}\frac{d}{dr}\left(r\frac{dz(r)}{dr}\right)\right]=p,\\
z(r_0)=0,\frac{dz(r)}{dr}\rvert_{r=r_0}=0,
\end{array}
\label{}
\end{equation}
the differential equation can be solved analytically to obtain equation \ref{eq:blistershape} with blister radius, $r_0$, and a maximum deflection given by: 
\begin{equation}
z_0=\frac{pr_0^4}{64D}.
\label{eq:z0r0purebending}
\end{equation}
The above equation relates the blister shape to the internal pressure of the blister but only takes into account the bending moment of the blister. This means that for small deflections $z_0<<t$ the blister height scales linearly with pressure. As will be shown in the next section, a correction due to stretching should be taken into account for large deflections. The analytical solution presented in equation \ref{eq:blistershape} is fitted in figure \ref{fig:AFMprofile} (solid lines) to experimentally measured AFM profiles of Mo/Si blister caps. To fit equation \ref{eq:blistershape}, $z_0$ and $r_0$ are taken as free parameters. As can be seen in figure \ref{fig:AFMprofile}, there is a good fit between the analytical shape function and the measured AFM profiles of the blister cap. The residual of the fit as given in the bottom graph is typically less than 8\%. The fit of equation \ref{eq:blistershape} overestimates the measured blister radius, which can be seen by the increase in the residual near the edge of the blister.  

\subsection{\label{subsec:energybalance}Energy balance of blister cap}

Several other calculations for the energy of the blister cap can be found in literature. For example \citet{selvadurai2007} balance the elastic strain energy including substrate deformation by the surface energy to determine the adhesion energy of thin films, while \citet{freund1997} and \citet{hong2006} take only the strain energy and surface energy into account to determine the minimal ion dose required for blisters to form. 

For the blisters formed in a Mo/Si multilayer the following equation is used to calculate the total potential energy of the blister cap:
\begin{equation}
E_{tot}(z_0,r_0)=U_b+U_s+\Gamma+W_{exp}+E_{int}.
\label{eq:energybalance}
\end{equation} 
In which the surface energy, $\Gamma$, the elastic bending energy, $U_b$, and the stretching energy, $U_s$, is balanced by the expansion work done by the blister’s internal pressure, $W_{exp}$, and the release of intrinsic compressive stress energy, $E_{int}$. In this calculation, the strain energy of the substrate is neglected as no significant deformation is observed in TEM images (see for example figure \ref{fig:TEMimage}). To calculate $U_b$ and $U_s$, the blister cap is assumed to be isotropic and elastic with a shape given by equation \ref{eq:blistershape}.

For small deflections ($z_0<r_0$) of a thin film, the bending energy in cylindrical coordinates is defined as\cite{timoshenko1987}:
\begin{eqnarray}
U_b=&&\frac{1}{2}D\nonumber\iint\left\{\left(\frac{\partial^2 z}{\partial r^2}+\frac{1}{r}\frac{\partial z}{\partial r}\right)^2-\right.\nonumber\\
&&\left.2(1-\nu)\frac{\partial^2 z}{\partial r^2}\left(\frac{1}{r}\frac{\partial^2 z}{\partial r^2}\right)\right\}rdrd\theta=\nonumber\\
&&\frac{32}{3}\pi D\left(\frac{z_0}{r_0}\right)^2.
\label{eq:bendingenergy}
\end{eqnarray}

If the deflection of the blister cap becomes comparable to the thickness, $z_0>t$, the stretching term, $U_s$, becomes significant. In this case, the in plane radial displacement $u(r)$ must be taken into account. Following the procedure of virtual displacement from \citet{timoshenko1987}, and taking for the radial displacement:
\begin{equation}
u(r)=r(r_0-r)(C_1+C_2r),
\label{}
\end{equation} 
the corresponding stretching energy is given by:
\begin{eqnarray}
U_s=\frac{\pi Et}{1-\nu^2}\int_{0}^{r_0}(\varepsilon_r^2+\varepsilon_\theta^2+2\varepsilon_r\varepsilon_\theta)rdr\nonumber\\
\varepsilon_t=\frac{\partial u}{\partial r}+\frac{1}{2}\left(\frac{\partial z}{\partial r}\right)^2,\hspace{2mm}\varepsilon_\theta=\frac{u}{r}.
\label{eq:stretchingenergy1}
\end{eqnarray}
By minimizing the stretching energy, constants $C_1$ and $C_2$ can be calculated by taking the partial derivatives $(\partial U_s)/(\partial C_1)=(\partial U_s)/(\partial C_2 )=0$. This reduces equation \ref{eq:stretchingenergy1} to:
\begin{subequations}
	\label{eq:stretchingenergy2}
	\begin{equation}
	U_s=\frac{32}{3}\pi D\left(\frac{z_0}{r_0}\right)^2\left\{\frac{3}{32}C\left(\frac{z_0}{t}\right)^2\right\}
	\label{}
	\end{equation}
	\begin{equation}
	C=\frac{-5585\nu^2+8500\nu+15010}{6615}.
	\label{}
	\end{equation}
\end{subequations}

The surface energy released by the blister is given by the delaminated area as: 
\begin{equation}
\Gamma=2\gamma\pi r_0^2.
\label{eq:surfaceenergy}
\end{equation}
with $\gamma$ the surface energy of the blister cap to substrate interface.

The work done by isothermally expanding $n$ gas particles inside a blister cavity in terms of pressure and volume is:
\begin{equation}
W_{exp}=-\int_{V_0}^{V_1}p_{EOS}(V)dV+W_0.
\label{eq:expansionwork1}
\end{equation}
Where $p_{EOS}(V)$ is the pressure as a function of volume, which is given by the equation of state (EOS). For large blisters the EOS is simply the ideal gas law, but as the blister volume approaches zero, the gas significantly deviates from the ideal gas law due to particle interactions. Around the stationary point of the blister, the following EOS of state can be used, as found experimentally by \citet{michels1959} for H$_2$ pressures in the range of 2-300 MPa \cite{michels1959}:
\begin{eqnarray}
p_{EOS}(V)=&&A\frac{n}{V}\left\{1+B\frac{n}{V}+C\frac{n^2}{V^2}+D\frac{n^3}{V^3}+E\frac{n^4}{V^4}+\right.\nonumber\\
&&\left.F\frac{n^5}{V^5}\right\},
\label{eq:EOS}
\end{eqnarray}
with $n$ the number of particles in moles, $V$ the volume in cubic meter and coefficients $A$ through $F$ as given in table \ref{tab:EOScoefficients}. With the above EOS, the molar density approaches that of solid hydrogen for pressures around 300 MPa.
\begin{table}
	\caption{\label{tab:EOScoefficients}Coefficients for Equation of State (EOS) given be equation \ref{eq:EOS} for a temperature T=298K}
	\begin{ruledtabular}
		\begin{tabular}{clccl}
		&Value&&&Value\\
		\hline
		A(=RT)	&$2479.62$ 				&&D& $3.3804\times10^{-15}$\\
		B		&$1.4384\times10^{-5}$	&&E& $9.2492\times10^{-20}$\\
		C		&$3.5637\times10^{-10}$ &&F& $-4.7594\times10^{-25}$\\
		\end{tabular}
	\end{ruledtabular}
\end{table}
For the expansion work, this leads to the equation:
\begin{subequations}
	\label{eq:expansionwork2}
	\begin{equation}
	W_{exp}=W(V_1)-W(V_0)+W_0
	\end{equation}
	\begin{equation}
	W(V)=-An\left\{ln(V)-B\frac{n}{V}-\frac{1}{2}C\frac{n^2}{V^2}+...\right\}\\
	\end{equation}
	\begin{equation}
	V=\frac{1}{3}\pi r_0^2z_0, 
	\end{equation}
\end{subequations}
where $W_0$ is the expansion work done for pressures above 300 MPa, and $W(V_1)-W(V_0)$ is the expansion work done for pressures within the validity range of the EOS. We assume that $W_0$ is constant for all blisters formed. The blister volume is calculated by taking the volume integral of equation \ref{eq:blistershape}.

Depending on the deposition process of the multilayer, the average stress of the Mo/Si bilayer can vary from hundreds of MPa pressure compressive to tensile \cite{windt1995}. For an initially compressively stressed blister cap the energy released by the delaminated layer is given by: 
\begin{equation}
E_{int}=-\frac{1-\nu}{E}\sigma_{int}^2t\pi r_0^2,
\label{eq:intrinsicenergy}
\end{equation}
Where $\sigma_{int}$ is the average compressive stress in the thin layer. If the film has a tensile stress the sign of the energy is changed and additional energy needs to be added to deflect the surface outward. One can show that for a deposited multilayer with material parameters as shown in table \ref{tab:defaultparameters}, the intrinsic stress has only a minor effect on the energy balance. (Compared to $U_b$ and $U_s$, the intrinsic stress is about two orders of magnitude lower,  $\approx10^{-14}$J compared to $\approx10^{-16}$J).   

\begin{figure}
	\includegraphics[width=1\columnwidth]{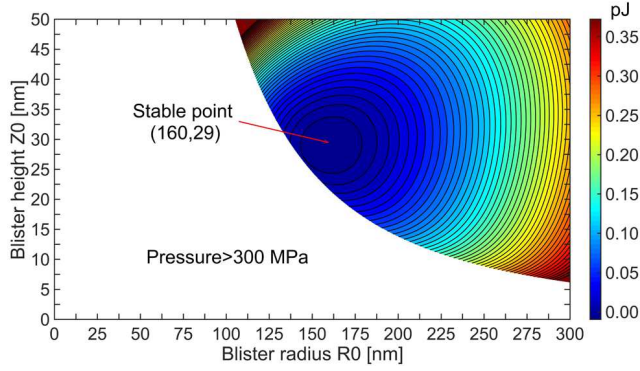}
	\caption{\label{fig:Energysurface}Contour plot of $E_{tot}(z_0,r_0)$ for 15 million trapped hydrogen particles. At $(160,29)$ a stable minimum is found for the total energy of the blister cap}
\end{figure}

Adding all energy terms as given by equation \ref{eq:energybalance}, the total energy of the blister cap as a function of blister radius and height can be found for a fixed number of $n$ hydrogen particles inside the blister cavity. For a Mo/Si multilayer the contour lines of the energy surface for 15 million trapped particles is shown in figure \ref{fig:Energysurface}. In this calculation material constants and dimensions are used as shown in table \ref{tab:defaultparameters}. 
\begin{table}
	\caption{\label{tab:defaultparameters}Material constants and dimensions used to model the blister cap in a Mo/Si multilayer}
	\begin{ruledtabular}
		\begin{tabular}{cccc}
			&Parameter&Value&\\
			\hline
			&$E$		&215 GPa\footnotemark[1]&\\
			&$\nu$		&0.18\footnotemark[1]&\\
			&$\gamma$	&1.05 J/m$^2$ &\\
			&$\sigma_{int}$&500 MPa &\\
			&$t$		&7 nm&
		\end{tabular}
	\end{ruledtabular}
\footnotetext[1]{Calculated values taken from \citet{loopstra1991}.}
\end{table}

In the white area the pressure inside the blister exceeds 300 MPa. At those pressure the hydrogen densities approaches that of solid hydrogen and the expansion work can no longer be calculated from the EOS. For the limiting case it can be seen that as the volume goes to zero, the expansion work tends to infinity. On the other hand, the surface energy (blister radius) and stretching energy (blister height) will increase continuously for an increasing blister size. Thus, for a fixed number of trapped molecular hydrogen inside the blister, a stable minimum in the blister cap energy can be found, as indicated by the red arrow in figure \ref{fig:Energysurface}.

\subsection{\label{subsec:stableblistershape}Stable blister shape}
\begin{figure}
	\includegraphics[width=1\columnwidth]{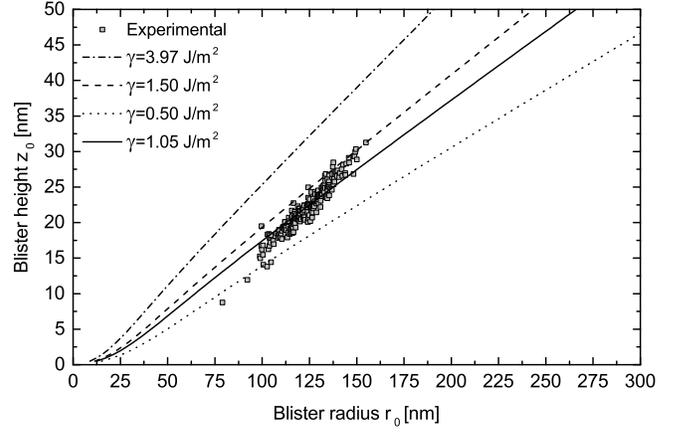}
	\caption{\label{fig:blisterequilibrium}Equilibrium of blister radius and height as a function for four different surface energies. The experimental data is an AFM measurement of blisters formed on a Mo/Si multilayer after hydrogen exposure.}
\end{figure}

To find the stable point as shown in figure \ref{fig:Energysurface} the partial derivatives of $E_{tot}$ with respect to $r_0$ and $z_0$ are taken. This leads to the following equations for the stable point:
\begin{subequations}
	\label{eq:stableblister}
	\begin{equation}
	z_{0,eq}=\frac{p_{EOS}(n,r_{0,eq},z_{0,eq})r_{0,eq}^4}{64D}\frac{1}{1+\frac{3}{16}C\left(\frac{z_{0,eq}}{t}\right)^2}
	\label{subeq:blisterpressure}
	\end{equation}
	\begin{equation}
	\nonumber
	\end{equation}
	\begin{equation}
	r_{0,eq}=\sqrt[4]{
		\frac{16Dz_{0,eq}^2}{\gamma-\frac{1-\nu}{2E}\sigma_{int}^2t}
		\left\{
		1+\frac{5}{32}C\left(
			\frac{z_{0,eq}}{t}
			\right)^2
		\right\}
		}.
	\label{subeq:stableblister}
	\end{equation}
\end{subequations}
The first equation relates the blister’s internal pressure to its dimensions $r_0$ and $z_0$. It is comparable with equation \ref{eq:z0r0purebending} but an additional term is included that takes the stretching of the blister cap into account. With increasing number of particles the stable blister size increases. The second equation gives the minimum in blister cap energy surface. If $z_{0,eq}\gg t$ there is a linear dependence between the blister radius and blister height. In figure \ref{fig:blisterequilibrium} the stable blister dimensions for four different surface energies are calculated taken the values as given in table \ref{tab:defaultparameters}. It can be seen that for increasing surface energies the ratio between blister height and radius increase. To verify the model, data is taken from an atomic force microscope measurement on a blistered Mo/Si multilayer surface being exposed to hydrogen. From the graph it is seen that the surface energy of the delaminated surface is around 1.05 J/m$^2$ which equals the surface energy of a-Si as can be found in literature: 1.05$\pm$0.14 J/m$^2$\hspace{1mm} \cite{hara2005}. For comparison the surface energy of (001) Mo and (001) MoSi$_2$ is around 3.97 J/m$^2$ and 3.86 J/m$^2$ respectively \cite{hong1992}. So based on the model it is expected that the delamination has taken place within the a-Si layer.

\subsection{\label{subsec:hydrogendensity}Hydrogen density and pressure inside a Mo/Si multilayer blister}
\begin{figure}
	\includegraphics[width=1\columnwidth]{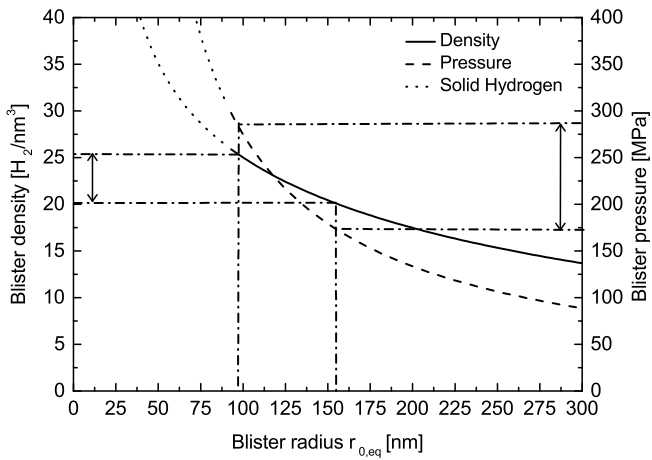}
	\caption{\label{fig:blisterpressure}Blister pressure and density as a function of the stable blister radius. The dash dotted line gives the experimental measured blister radii and their corresponding ranges for the pressure and density.}
\end{figure}

From a measured blister shape, the hydrogen density and pressure inside the blister can be calculated with equations \ref{eq:stableblister} and \ref{eq:EOS}. In figure \ref{fig:blisterpressure} the blister pressure and density is given as a function of the stable blister radius. For blister radii smaller than $\approx$90 nm the hydrogen density necessary for stable blisters to form approach values of solid hydrogen (dotted line). The dash dotted line gives the measured blister radius range and the corresponding range in blister density (20.1-25.3 H$_2$/nm$^3$) and pressure (175-280 MPa). 

\begin{figure}
	\includegraphics[width=1\columnwidth]{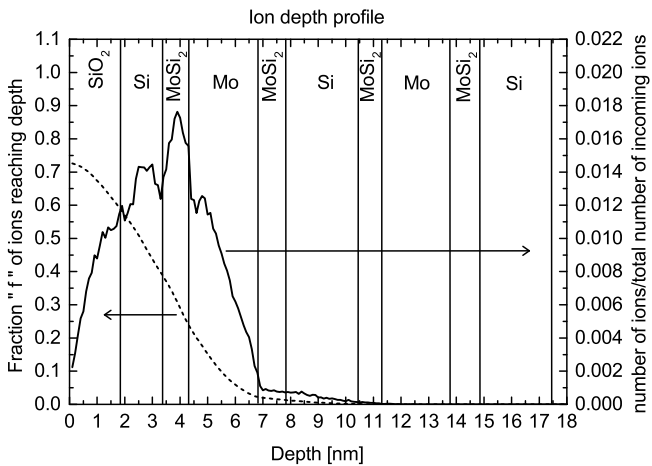}
	\caption{\label{fig:SRIM}SRIM calculation of hydrogen ion penetration depth in a Mo/Si multilayer with native SiO$_2$ on top. Fraction $f$ of total flux that reaches a certain depth (dashed line) and number of ions at certain depth (solid line) are given.}
\end{figure}

When the stable blister shape and density are known, an estimate can be made on the minimum dose required to form the blister. For an observed blister with a radius of 98 nm and corresponding height of 17 nm, the local hydrogen density has to be 25.3 H$_2$/nm$^3$. This means that $N_{min}$=4.3 million particles have to get trapped in the blister cavity. The number of hydrogen particles reaching the blister volume $N_{min}$ is given by: 
\begin{equation}
N_{min}=\frac{1}{2}fd\pi r_0^2.
\label{eq:minimumdose}
\end{equation}
In this equation, $d$ is the incident hydrogen ion dose per unit area and $f$ the fraction of the incoming ions that can penetrate through the blister cap. The factor of one half takes into account recombination of hydrogen ions to stable molecular hydrogen. In this equation the diffusion of hydrogen after implantation is neglected. To estimate $f$, an SRIM calculation was performed for a Mo/Si multilayer irradiated by 100 eV hydrogen ions (see figure \ref{fig:SRIM}) \cite{ziegler1985,ziegler2016}. The fraction of the total flux that can penetrate through the first bilayer is $f\approx6.8\times10^{-3}$. Filling in those number in equation \ref{eq:minimumdose} gives a minimum required ion dose of $d=4.2\times10^{18}$ ions/cm$^2$. This is indeed below the actual measurement dose of $1.25\times10^{19}$ ions/cm$^2$. 

\subsection{\label{subsec:blisterstabilization}Blister stabilization}
In the analysis above only the static case of the blister is considered where the number of trapped hydrogen particles inside the blister is fixed. But in general, depending on the in- and outflux of hydrogen ($H_{in}$, $H_{out}$), three cases can be distinguished: i) $H_{in}>H_{out}$: the number of trapped hydrogen particles is increasing and the blister grows, ii) $H_{in}=H_{out}$: the number of trapped hydrogen particles is fixed and the blister is stable at its energetically most favorable shape, iii) $H_{in}<H_{out}$: the number of trapped hydrogen particles is decreasing and the blister size decreases assuming the deformation is completely elastic. By knowing the in- and outflux as function of time the dynamic behavior of the blister can be described.   

The influx of hydrogen per unit of time  $H_{in}$ H$_2$/s is given by:
\begin{equation}
H_{in}=\frac{1}{2}\phi f\pi r_0^2,
\label{eq:influx}
\end{equation}
with $\phi$ ions/cm$^2$s the hydrogen ion flux at the surface and $f$ the fraction of ions that can penetrate through the blister cap. The out diffusion, $H_{out}$ H$_2$/s, of hydrogen can be estimated using Fick’s law:
\begin{equation}
H_{out}=D_{H_2}\frac{n/V}{t}S_{blister}.
\label{eq:outflux}
\end{equation}
The diffusion constant $D_{H_2}$ cm$^2$/s depends on the blister cap material and $n$ is the number of hydrogen particles that are trapped inside the blister with a volume $V$. $S_{blister}$ is the blister cap surface area:
\begin{subequations}
	\label{eq:blistersurface}
	\begin{equation}
	S_{blister}=\iint z(r)dS\approx\pi r_0^2F.
	\label{}
	\end{equation}
	\begin{equation}
	F=1-0.0036\left(\frac{z_0}{r_0}\right)+0.715\left(\frac{z_0}{r_0}\right)^2-0.205\left(\frac{z_0}{r_0}\right)^3
	\label{}
	\end{equation}
\end{subequations}

From the final shape of the blister, the diffusion constant $D_{H_2}$ can be calculated by combining equation \ref{eq:influx} and \ref{eq:outflux}:
\begin{equation}
D_{H_2}=\frac{\phi ftV}{2n}\frac{1}{F}.
\label{eq:fluxbalance}
\end{equation}
Taking the blister shape of the measured data with a number density between 20.1-25.3 H$_2$/nm$^3$ exposed at a constant flux of $\phi=1\times10^{14}$ ions/cm$^2$s, the diffusion constant is around $D_{H_2}=(10\pm1)\times10^{-18}$ cm$^2$/s. This is about an order of magnitude higher than the literature value for hydrogen diffusion in c-Si\cite{lide2005}: $D_{H_2}=2.36\times10^{-18}$ cm$^2$/s.   

Before the blister is stabilized the blister is growing what means that $H_{in}>H_{out}$. Because the influx of hydrogen particles per unit area is constant during the experiment, the outflux of hydrogen per unit area has to increase during the blister growth until the influx is balanced and the blister growth will stop. This means that regarding equation \ref{eq:outflux} that the hydrogen density, n/V, and/or the diffusion constant, $D_{H_2}$, has to increase in time. At the start of the hydrogen exposure the concentration inside the multilayer is zero and it is slowly increasing with time. While the influx is set instantaneously  by the ion flux and energy, the outflux is increasing towards the influx value with a certain delay depending on the diffusion constant $D_{H_2}$ and thickness of the blister cap. The timescale for out diffusion can be estimated with:
\begin{equation}
\tau=\frac{t^2}{D_{H_2}}\approx10^4s.
\label{eq:diffusiontimescale}
\end{equation}
This is the typical time it takes for a hydrogen particle to diffuse through the blister cap. During this time about $\tau\times H_{in}=$ 4 million H$_2$ particles have penetrated through the blister cap. This is about the number of particles needed to form a stable blister of 100 nm in radius. For a large diffusion constant the delay between in and outflux becomes smaller and no blisters will form at all as not sufficient hydrogen can get trapped. On the other hand a lower diffusion coefficient will increase the blister size as more particles get trapped before a balance between in and outflux is established. 

Although it is possible to form a stable blister where fluxes are balanced, this point doesn't have good stability. As the blister radius increases beyond the stable point, the influx of hydrogen per unit area becomes larger than the outflux and the blister can continue to grow until it bursts. This contradicts the stable blister shape observed on a Mo/Si multilayer. A possible explanation for this can be found in a time dependent diffusion constant. Studies on the hydrogenation of amorphous silicon have shown that the permeability of a-Si can change significantly during the hydrogenation \cite{beyer2016}. These studies suggest that before the Si layer is fully saturated, only atomic hydrogen can diffuse through the silicon layer, and the rate of diffusion is rather low. However, once a-Si:H is fully hydrogenated, $H_2$ can freely and rapidly diffuse through the layer. This means that during the blister formation process this enhanced outdiffusion of hydrogen may prevent further blister growth. If a certain change in the diffusion is sufficient to stop the blister from growing depends on the time and scale of the change. If the change in diffusion is around the initiation point of the blister, a doubling in the diffusion constant could be enough to stop the blister growth process. In addition, \citet{coupeau2013} also suggest that changes in the diffusion process occur during blister formation. \citeauthor{coupeau2013} show that the evolution of blisters on a pure silicon surface has small discontinuities over time, which the authors address to changes in the diffusion process.

\section{\label{sec:conclusion}conclusion}
The results presented here show that an elastic energy balance of a blister cap can be applied to hydrogen induced blister formation in Mo/Si multilayer mirrors. The model shows that the blister formation is mainly caused by the accumulation of hydrogen under the blister cap and a that a stable blister can be formed with a fixed number of particles inside. From the measured blister radius and height the surface energy, pressure and minimum hydrogen dose for blister formation could be calculated from the model. In second part the diffusion and penetration through the blister cap was considered. Given a fixed influx of hydrogen and assuming a linear outdiffusion of hydrogen, a diffusion constant for hydrogen through the blister cap could be calculated. Furthermore, the lack of blister expansion once the a stable blister has been formed is likely caused by a time dependent permeability of the blister cap.

\begin{acknowledgments}
The authors would like to thank Dimitry Lopaev and Slava Medvedev for their contributions to the scientific discussions about this model. This work is part of the research programme CP3E (Controlling photon and plasma induced processes at EUV optical surfaces) of FOM (Stichting voor Fundamenteel Onderzoek der Materie) with the financial support of NWO (Nederlandse organisatie voor Wetenschappelijk Onderzoek). The CP3E programme is co-financed by Carl Zeiss SMT and ASML, and AgentschapNL through the EXEPT programme.
\end{acknowledgments}

\bibliography{RAJMvdBos_blistermodel_v2sep16}

\begin{thebibliography}{24}%
\makeatletter
\providecommand \@ifxundefined [1]{%
 \@ifx{#1\undefined}
}%
\providecommand \@ifnum [1]{%
 \ifnum #1\expandafter \@firstoftwo
 \else \expandafter \@secondoftwo
 \fi
}%
\providecommand \@ifx [1]{%
 \ifx #1\expandafter \@firstoftwo
 \else \expandafter \@secondoftwo
 \fi
}%
\providecommand \natexlab [1]{#1}%
\providecommand \enquote  [1]{``#1''}%
\providecommand \bibnamefont  [1]{#1}%
\providecommand \bibfnamefont [1]{#1}%
\providecommand \citenamefont [1]{#1}%
\providecommand \href@noop [0]{\@secondoftwo}%
\providecommand \href [0]{\begingroup \@sanitize@url \@href}%
\providecommand \@href[1]{\@@startlink{#1}\@@href}%
\providecommand \@@href[1]{\endgroup#1\@@endlink}%
\providecommand \@sanitize@url [0]{\catcode `\\12\catcode `\$12\catcode
  `\&12\catcode `\#12\catcode `\^12\catcode `\_12\catcode `\%12\relax}%
\providecommand \@@startlink[1]{}%
\providecommand \@@endlink[0]{}%
\providecommand \url  [0]{\begingroup\@sanitize@url \@url }%
\providecommand \@url [1]{\endgroup\@href {#1}{\urlprefix }}%
\providecommand \urlprefix  [0]{URL }%
\providecommand \Eprint [0]{\href }%
\providecommand \doibase [0]{http://dx.doi.org/}%
\providecommand \selectlanguage [0]{\@gobble}%
\providecommand \bibinfo  [0]{\@secondoftwo}%
\providecommand \bibfield  [0]{\@secondoftwo}%
\providecommand \translation [1]{[#1]}%
\providecommand \BibitemOpen [0]{}%
\providecommand \bibitemStop [0]{}%
\providecommand \bibitemNoStop [0]{.\EOS\space}%
\providecommand \EOS [0]{\spacefactor3000\relax}%
\providecommand \BibitemShut  [1]{\csname bibitem#1\endcsname}%
\let\auto@bib@innerbib\@empty
%</preamble>
\bibitem [{\citenamefont {Pelizzo}\ \emph {et~al.}(2011)\citenamefont
  {Pelizzo}, \citenamefont {Corso}, \citenamefont {Zuppella}, \citenamefont
  {Windt}, \citenamefont {Mattei},\ and\ \citenamefont
  {Nicolosi}}]{pelizzo2011}%
  \BibitemOpen
  \bibfield  {author} {\bibinfo {author} {\bibfnamefont {M.~G.}\ \bibnamefont
  {Pelizzo}}, \bibinfo {author} {\bibfnamefont {A.~J.}\ \bibnamefont {Corso}},
  \bibinfo {author} {\bibfnamefont {P.}~\bibnamefont {Zuppella}}, \bibinfo
  {author} {\bibfnamefont {D.~L.}\ \bibnamefont {Windt}}, \bibinfo {author}
  {\bibfnamefont {G.}~\bibnamefont {Mattei}}, \ and\ \bibinfo {author}
  {\bibfnamefont {P.}~\bibnamefont {Nicolosi}},\ }\href {\doibase
  10.1364/OE.19.014838} {\bibfield  {journal} {\bibinfo  {journal} {Optics
  express}\ }\textbf {\bibinfo {volume} {19}},\ \bibinfo {pages} {14838}
  (\bibinfo {year} {2011})}\BibitemShut {NoStop}%
\bibitem [{\citenamefont {Teyssier}\ \emph {et~al.}(2002)\citenamefont
  {Teyssier}, \citenamefont {Quesnel}, \citenamefont {Muffato},\ and\
  \citenamefont {Schiavone}}]{teyssier2002}%
  \BibitemOpen
  \bibfield  {author} {\bibinfo {author} {\bibfnamefont {C.}~\bibnamefont
  {Teyssier}}, \bibinfo {author} {\bibfnamefont {E.}~\bibnamefont {Quesnel}},
  \bibinfo {author} {\bibfnamefont {V.}~\bibnamefont {Muffato}}, \ and\
  \bibinfo {author} {\bibfnamefont {P.}~\bibnamefont {Schiavone}},\ }\href
  {\doibase 10.1016/S0167-9317(02)00506-3} {\bibfield  {journal} {\bibinfo
  {journal} {Microelectronic Engineering}\ }\textbf {\bibinfo {volume}
  {61-62}},\ \bibinfo {pages} {241} (\bibinfo {year} {2002})}\BibitemShut
  {NoStop}%
\bibitem [{\citenamefont {Evans}(1976)}]{evans1976}%
  \BibitemOpen
  \bibfield  {author} {\bibinfo {author} {\bibfnamefont {J.~H.}\ \bibnamefont
  {Evans}},\ }\href {\doibase 10.1016/0022-3115(76)90092-1} {\bibfield
  {journal} {\bibinfo  {journal} {Journal of Nuclear Materials}\ }\textbf
  {\bibinfo {volume} {61}},\ \bibinfo {pages} {1} (\bibinfo {year}
  {1976})}\BibitemShut {NoStop}%
\bibitem [{\citenamefont {Feng}\ and\ \citenamefont {Huang}(2004)}]{feng2004}%
  \BibitemOpen
  \bibfield  {author} {\bibinfo {author} {\bibfnamefont {X.~Q.}\ \bibnamefont
  {Feng}}\ and\ \bibinfo {author} {\bibfnamefont {Y.}~\bibnamefont {Huang}},\
  }\href {\doibase 10.1016/j.ijsolstr.2004.02.054} {\bibfield  {journal}
  {\bibinfo  {journal} {International Journal of Solids and Structures}\
  }\textbf {\bibinfo {volume} {41}},\ \bibinfo {pages} {4299} (\bibinfo {year}
  {2004})}\BibitemShut {NoStop}%
\bibitem [{\citenamefont {H{\"o}chbauer}\ \emph {et~al.}(2005)\citenamefont
  {H{\"o}chbauer}, \citenamefont {Misra}, \citenamefont {Hattar},\ and\
  \citenamefont {Hoagland}}]{hochbauer2005}%
  \BibitemOpen
  \bibfield  {author} {\bibinfo {author} {\bibfnamefont {T.}~\bibnamefont
  {H{\"o}chbauer}}, \bibinfo {author} {\bibfnamefont {A.}~\bibnamefont
  {Misra}}, \bibinfo {author} {\bibfnamefont {K.}~\bibnamefont {Hattar}}, \
  and\ \bibinfo {author} {\bibfnamefont {R.~G.}\ \bibnamefont {Hoagland}},\
  }\href {\doibase 10.1063/1.2149168} {\bibfield  {journal} {\bibinfo
  {journal} {Journal of Applied Physics}\ }\textbf {\bibinfo {volume} {98}},\
  \bibinfo {pages} {123516} (\bibinfo {year} {2005})}\BibitemShut {NoStop}%
\bibitem [{\citenamefont {Sharafat}\ \emph {et~al.}(2009)\citenamefont
  {Sharafat}, \citenamefont {Takahashi}, \citenamefont {Nagasawa},\ and\
  \citenamefont {Ghoniem}}]{sharafat2009}%
  \BibitemOpen
  \bibfield  {author} {\bibinfo {author} {\bibfnamefont {S.}~\bibnamefont
  {Sharafat}}, \bibinfo {author} {\bibfnamefont {A.}~\bibnamefont {Takahashi}},
  \bibinfo {author} {\bibfnamefont {K.}~\bibnamefont {Nagasawa}}, \ and\
  \bibinfo {author} {\bibfnamefont {N.}~\bibnamefont {Ghoniem}},\ }\href
  {\doibase 10.1016/j.jnucmat.2009.02.027} {\bibfield  {journal} {\bibinfo
  {journal} {Journal of Nuclear Materials}\ }\textbf {\bibinfo {volume}
  {389}},\ \bibinfo {pages} {203} (\bibinfo {year} {2009})}\BibitemShut
  {NoStop}%
\bibitem [{\citenamefont {Xie}\ \emph {et~al.}(2015)\citenamefont {Xie},
  \citenamefont {Wang}, \citenamefont {Sun}, \citenamefont {Li}, \citenamefont
  {Ma},\ and\ \citenamefont {Shan}}]{xie2015}%
  \BibitemOpen
  \bibfield  {author} {\bibinfo {author} {\bibfnamefont {D.~G.}\ \bibnamefont
  {Xie}}, \bibinfo {author} {\bibfnamefont {Z.~J.}\ \bibnamefont {Wang}},
  \bibinfo {author} {\bibfnamefont {J.}~\bibnamefont {Sun}}, \bibinfo {author}
  {\bibfnamefont {J.}~\bibnamefont {Li}}, \bibinfo {author} {\bibfnamefont
  {E.}~\bibnamefont {Ma}}, \ and\ \bibinfo {author} {\bibfnamefont {Z.~W.}\
  \bibnamefont {Shan}},\ }\href {\doibase 10.1038/nmat4336} {\bibfield
  {journal} {\bibinfo  {journal} {Nature Materials}\ }\textbf {\bibinfo
  {volume} {14}},\ \bibinfo {pages} {899} (\bibinfo {year} {2015})}\BibitemShut
  {NoStop}%
\bibitem [{\citenamefont {Johnson}, \citenamefont {Thomson},\ and\
  \citenamefont {Reader}(1999)}]{johnson1999}%
  \BibitemOpen
  \bibfield  {author} {\bibinfo {author} {\bibfnamefont {P.~B.}\ \bibnamefont
  {Johnson}}, \bibinfo {author} {\bibfnamefont {R.~W.}\ \bibnamefont
  {Thomson}}, \ and\ \bibinfo {author} {\bibfnamefont {K.}~\bibnamefont
  {Reader}},\ }\href {\doibase 10.1016/S0022-3115(99)00046-X} {\bibfield
  {journal} {\bibinfo  {journal} {Journal of Nuclear Materials}\ }\textbf
  {\bibinfo {volume} {273}},\ \bibinfo {pages} {117} (\bibinfo {year}
  {1999})}\BibitemShut {NoStop}%
\bibitem [{\citenamefont {Freund}(1997)}]{freund1997}%
  \BibitemOpen
  \bibfield  {author} {\bibinfo {author} {\bibfnamefont {L.~B.}\ \bibnamefont
  {Freund}},\ }\href {\doibase 10.1063/1.119219} {\bibfield  {journal}
  {\bibinfo  {journal} {Applied Physics Letters}\ }\textbf {\bibinfo {volume}
  {70}},\ \bibinfo {pages} {3519} (\bibinfo {year} {1997})}\BibitemShut
  {NoStop}%
\bibitem [{\citenamefont {Hong}\ and\ \citenamefont {Cheong}(2006)}]{hong2006}%
  \BibitemOpen
  \bibfield  {author} {\bibinfo {author} {\bibfnamefont {J.~W.}\ \bibnamefont
  {Hong}}\ and\ \bibinfo {author} {\bibfnamefont {S.}~\bibnamefont {Cheong}},\
  }\href {\doibase 10.1063/1.2364040} {\bibfield  {journal} {\bibinfo
  {journal} {Journal of Applied Physics}\ }\textbf {\bibinfo {volume} {100}},\
  \bibinfo {pages} {094322} (\bibinfo {year} {2006})}\BibitemShut {NoStop}%
\bibitem [{\citenamefont {Mitani}\ and\ \citenamefont
  {G{\"o}sele}(1992)}]{mitani1992}%
  \BibitemOpen
  \bibfield  {author} {\bibinfo {author} {\bibfnamefont {K.}~\bibnamefont
  {Mitani}}\ and\ \bibinfo {author} {\bibfnamefont {U.~M.}\ \bibnamefont
  {G{\"o}sele}},\ }\href {\doibase 10.1007/BF00324337} {\bibfield  {journal}
  {\bibinfo  {journal} {Applied Physics A: Solids and Surfaces}\ }\textbf
  {\bibinfo {volume} {54}},\ \bibinfo {pages} {543} (\bibinfo {year}
  {1992})}\BibitemShut {NoStop}%
\bibitem [{\citenamefont {Selvadurai}(2007)}]{selvadurai2007}%
  \BibitemOpen
  \bibfield  {author} {\bibinfo {author} {\bibfnamefont {A.~P.~S.}\
  \bibnamefont {Selvadurai}},\ }\href {\doibase 10.1016/j.actamat.2007.04.038}
  {\bibfield  {journal} {\bibinfo  {journal} {Acta Materialia}\ }\textbf
  {\bibinfo {volume} {55}},\ \bibinfo {pages} {4679} (\bibinfo {year}
  {2007})}\BibitemShut {NoStop}%
\bibitem [{\citenamefont {Terreault}(2007)}]{terreault2007}%
  \BibitemOpen
  \bibfield  {author} {\bibinfo {author} {\bibfnamefont {B.}~\bibnamefont
  {Terreault}},\ }\href {\doibase 10.1002/pssa.200622520} {\bibfield  {journal}
  {\bibinfo  {journal} {{P}hysica {S}tatus {S}olidi A}\ }\textbf {\bibinfo
  {volume} {204}},\ \bibinfo {pages} {2129} (\bibinfo {year}
  {2007})}\BibitemShut {NoStop}%
\bibitem [{\citenamefont {Coupeau}\ \emph {et~al.}(2013)\citenamefont
  {Coupeau}, \citenamefont {Parry}, \citenamefont {Colin}, \citenamefont
  {David}, \citenamefont {Labanowski},\ and\ \citenamefont
  {Grilhé}}]{coupeau2013}%
  \BibitemOpen
  \bibfield  {author} {\bibinfo {author} {\bibfnamefont {C.}~\bibnamefont
  {Coupeau}}, \bibinfo {author} {\bibfnamefont {G.}~\bibnamefont {Parry}},
  \bibinfo {author} {\bibfnamefont {J.}~\bibnamefont {Colin}}, \bibinfo
  {author} {\bibfnamefont {M.~L.}\ \bibnamefont {David}}, \bibinfo {author}
  {\bibfnamefont {J.}~\bibnamefont {Labanowski}}, \ and\ \bibinfo {author}
  {\bibfnamefont {J.}~\bibnamefont {Grilhé}},\ }\href {\doibase
  10.1063/1.4813858} {\bibfield  {journal} {\bibinfo  {journal} {Applied
  Physics Letters}\ }\textbf {\bibinfo {volume} {103}},\ \bibinfo {pages}
  {031908} (\bibinfo {year} {2013})}\BibitemShut {NoStop}%
\bibitem [{\citenamefont {Timoshenko}\ and\ \citenamefont
  {Woinowsky-Krieger}(1959)}]{timoshenko1987}%
  \BibitemOpen
  \bibfield  {author} {\bibinfo {author} {\bibfnamefont {S.}~\bibnamefont
  {Timoshenko}}\ and\ \bibinfo {author} {\bibfnamefont {S.}~\bibnamefont
  {Woinowsky-Krieger}},\ }\href@noop {} {\emph {\bibinfo {title} {Theory of
  plates and shells}}},\ \bibinfo {edition} {2nd}\ ed.\ (\bibinfo  {publisher}
  {McGraw-Hill},\ \bibinfo {year} {1959})\BibitemShut {NoStop}%
\bibitem [{\citenamefont {Michels}\ \emph {et~al.}(1959)\citenamefont
  {Michels}, \citenamefont {de~Graaff}, \citenamefont {Wassenaar},
  \citenamefont {Levelt},\ and\ \citenamefont {Louwerse}}]{michels1959}%
  \BibitemOpen
  \bibfield  {author} {\bibinfo {author} {\bibfnamefont {A.}~\bibnamefont
  {Michels}}, \bibinfo {author} {\bibfnamefont {W.}~\bibnamefont {de~Graaff}},
  \bibinfo {author} {\bibfnamefont {T.}~\bibnamefont {Wassenaar}}, \bibinfo
  {author} {\bibfnamefont {J.~M.~H.}\ \bibnamefont {Levelt}}, \ and\ \bibinfo
  {author} {\bibfnamefont {P.}~\bibnamefont {Louwerse}},\ }\href {\doibase
  10.1016/S0031-8914(59)90713-X} {\bibfield  {journal} {\bibinfo  {journal}
  {Physica}\ }\textbf {\bibinfo {volume} {25}},\ \bibinfo {pages} {25}
  (\bibinfo {year} {1959})}\BibitemShut {NoStop}%
\bibitem [{\citenamefont {Windt}\ \emph {et~al.}(1995)\citenamefont {Windt},
  \citenamefont {Brown}, \citenamefont {Volkert},\ and\ \citenamefont
  {Waskiewicz}}]{windt1995}%
  \BibitemOpen
  \bibfield  {author} {\bibinfo {author} {\bibfnamefont {D.~L.}\ \bibnamefont
  {Windt}}, \bibinfo {author} {\bibfnamefont {W.~L.}\ \bibnamefont {Brown}},
  \bibinfo {author} {\bibfnamefont {C.~A.}\ \bibnamefont {Volkert}}, \ and\
  \bibinfo {author} {\bibfnamefont {W.~K.}\ \bibnamefont {Waskiewicz}},\ }\href
  {\doibase 10.1063/1.360164} {\bibfield  {journal} {\bibinfo  {journal}
  {Journal of Applied Physics}\ }\textbf {\bibinfo {volume} {78}},\ \bibinfo
  {pages} {2423} (\bibinfo {year} {1995})}\BibitemShut {NoStop}%
\bibitem [{\citenamefont {Loopstra}\ \emph {et~al.}(1991)\citenamefont
  {Loopstra}, \citenamefont {van Snek}, \citenamefont {de~Keijser},\ and\
  \citenamefont {Mittemeijer}}]{loopstra1991}%
  \BibitemOpen
  \bibfield  {author} {\bibinfo {author} {\bibfnamefont {O.~B.}\ \bibnamefont
  {Loopstra}}, \bibinfo {author} {\bibfnamefont {E.~R.}\ \bibnamefont {van
  Snek}}, \bibinfo {author} {\bibfnamefont {T.~H.}\ \bibnamefont {de~Keijser}},
  \ and\ \bibinfo {author} {\bibfnamefont {E.~J.}\ \bibnamefont
  {Mittemeijer}},\ }\href {\doibase 10.1103/PhysRevB.44.13519} {\bibfield
  {journal} {\bibinfo  {journal} {Physical Review B}\ }\textbf {\bibinfo
  {volume} {44}},\ \bibinfo {pages} {13519} (\bibinfo {year}
  {1991})}\BibitemShut {NoStop}%
\bibitem [{\citenamefont {Hara}\ \emph {et~al.}(2005)\citenamefont {Hara},
  \citenamefont {Izumi}, \citenamefont {Kumagai},\ and\ \citenamefont
  {Sakai}}]{hara2005}%
  \BibitemOpen
  \bibfield  {author} {\bibinfo {author} {\bibfnamefont {S.}~\bibnamefont
  {Hara}}, \bibinfo {author} {\bibfnamefont {S.}~\bibnamefont {Izumi}},
  \bibinfo {author} {\bibfnamefont {T.}~\bibnamefont {Kumagai}}, \ and\
  \bibinfo {author} {\bibfnamefont {S.}~\bibnamefont {Sakai}},\ }\href
  {\doibase 10.1016/j.susc.2005.03.061} {\bibfield  {journal} {\bibinfo
  {journal} {Surface Science}\ }\textbf {\bibinfo {volume} {585}},\ \bibinfo
  {pages} {17} (\bibinfo {year} {2005})}\BibitemShut {NoStop}%
\bibitem [{\citenamefont {Hong}\ \emph {et~al.}(1992)\citenamefont {Hong},
  \citenamefont {Smith}, \citenamefont {Srolovitz}, \citenamefont {Gay},\ and\
  \citenamefont {Richter}}]{hong1992}%
  \BibitemOpen
  \bibfield  {author} {\bibinfo {author} {\bibfnamefont {T.}~\bibnamefont
  {Hong}}, \bibinfo {author} {\bibfnamefont {J.~R.}\ \bibnamefont {Smith}},
  \bibinfo {author} {\bibfnamefont {D.~J.}\ \bibnamefont {Srolovitz}}, \bibinfo
  {author} {\bibfnamefont {J.~G.}\ \bibnamefont {Gay}}, \ and\ \bibinfo
  {author} {\bibfnamefont {R.}~\bibnamefont {Richter}},\ }\href {\doibase
  10.1103/PhysRevB.45.8775} {\bibfield  {journal} {\bibinfo  {journal}
  {Physical Review B}\ }\textbf {\bibinfo {volume} {45}},\ \bibinfo {pages}
  {8775} (\bibinfo {year} {1992})}\BibitemShut {NoStop}%
\bibitem [{\citenamefont {Ziegler}, \citenamefont {Biersack},\ and\
  \citenamefont {Littmark}(1985)}]{ziegler1985}%
  \BibitemOpen
  \bibfield  {author} {\bibinfo {author} {\bibfnamefont {J.}~\bibnamefont
  {Ziegler}}, \bibinfo {author} {\bibfnamefont {J.}~\bibnamefont {Biersack}}, \
  and\ \bibinfo {author} {\bibfnamefont {U.}~\bibnamefont {Littmark}},\
  }\href@noop {} {\emph {\bibinfo {title} {The Stopping and Range of Ions in
  Matter}}}\ (\bibinfo  {publisher} {Pergamon},\ \bibinfo {address} {New
  York},\ \bibinfo {year} {1985})\BibitemShut {NoStop}%
\bibitem [{zie()}]{ziegler2016}%
  \BibitemOpen
  \href@noop {} {\enquote {\bibinfo {title} {{SRIM}},}\ }\bibinfo
  {howpublished} {http://www.srim.org/},\ \bibinfo {note} {consulted on July
  2016}\BibitemShut {NoStop}%
\bibitem [{\citenamefont {Sharma}(2005)}]{lide2005}%
  \BibitemOpen
  \bibfield  {author} {\bibinfo {author} {\bibfnamefont {B.~L.}\ \bibnamefont
  {Sharma}},\ }\enquote {\bibinfo {title} {{CRC} handbook of {C}hemistry and
  {P}hysics},}\ \ (\bibinfo  {publisher} {CRC Press},\ \bibinfo {year} {2005})\
  Chap.~\bibinfo {chapter} {12},\ \bibinfo {note} {internet version
  2005}\BibitemShut {NoStop}%
\bibitem [{\citenamefont {Beyer}(2016)}]{beyer2016}%
  \BibitemOpen
  \bibfield  {author} {\bibinfo {author} {\bibfnamefont {W.}~\bibnamefont
  {Beyer}},\ }\href {\doibase 10.1002/pssa.201532976} {\bibfield  {journal}
  {\bibinfo  {journal} {{P}hysica {S}tatus {S}olidi A}\ }\textbf {\bibinfo
  {volume} {213}},\ \bibinfo {pages} {1661} (\bibinfo {year}
  {2016})}\BibitemShut {NoStop}%
\end{thebibliography}%

\end{document}